\def\ga{\gamma}
\def\de{\delta}

\def\ka{\kappa}

\def\si{\sigma}

\def\ph{\phi}

\def\ps{\psi}

\def\Si{\Sigma}

\def\cL{{\cal L}}
\def\L{\textrm{L}}
\def\NR{\textrm{NR}}

\def\psb{\overline{\psi}}

\def\eff{\textrm{asy}}

\def\bt{{\widetilde b}}

\newcommand{\beq}{\begin{equation}}
\newcommand{\eeq}{\end{equation}}
\newcommand{\bea}{\begin{eqnarray}}
\newcommand{\eea}{\end{eqnarray}}

\newcommand{\nn}{\nonumber}

\documentclass{ws-procs9x6-cpt22}
\begin{document}

\newcommand{\refeq}[1]{(\ref{#1})}
\def\etal {{\it et al.}}

\title{Measuring Lorentz Violation
in Weak Gravity Fields}

\author{Zonghao Li}

\address{Department of Physics, Indiana University,
Bloomington, IN 47405, USA}

\begin{abstract}
Many new linearized coefficients for Lorentz violation
are discovered in our recent work
on the construction of a generic Lorentz-violating 
effective field theory in curved spacetime.
The new coefficients can be constrained 
by experiments in weak gravity fields.
In this work,
we compare experiments in different gravitational potentials
and study three types of gravity-related experiments:
free-fall, gravitational interferometer, 
and gravitational bound-state experiments.
First constraints on the new coefficients for Lorentz violation
are extracted from those experiments.
\end{abstract}

\bodymatter

\section{Lorentz violation in gravity}

In recent years,
Lorentz violation has been a popular topic
in the search for physics 
beyond the Standard Model (SM) and General Relativity (GR).
The Standard-Model Extension (SME)\cite{ck,ak04}
has been widely used as a comprehensive framework
to study Lorentz violation 
in the context of effective field theory.
The minimal terms in the Lagrange density 
of the SME in curved spacetime
were constructed by Kosteleck\'y in 2004,\cite{ak04}
and the nonminimal terms 
were systematically constructed in our recent work.\cite{kl19,kl21a}
The linearizations of those terms in weak gravity fields
were also obtained.\cite{kl21b}
The present contribution to the proceedings of CPT'22 
studies the experimental implications of the linearized terms 
with a focus on matter--gravity couplings
in weak gravity fields.
This work is based on the results in Ref.~[\refcite{kl21b}].

\section{Potential-dependent experiments}

An interesting implication of the linearized Lagrange density
constructed in our recent work\cite{kl21b}
is that the measured SME coefficients for Lorentz violation 
can depend on the gravitational potential of the laboratory.
Coefficients for Lorentz violation have been measured 
in many experiments 
under the assumption that spacetime is flat.\cite{datatables}
However,
these experiments are typically performed at different elevations
and hence at different gravitational potentials,
so the SME coefficients can depend on the potentials.

Taking the $b$-type coefficients as an example,
we know that the term in the Lagrange density 
containing the minimal $b^\ka$ coefficient
is $\cL\supset b^\ka \psb\ga_{\ka}\ps$
in flat spacetime.\cite{ck}
Adding couplings with the gravitational field,
we can write the generalization of the term 
in a weak gravity field as\cite{kl21b}
\beq
\cL\supset (b_\eff^\ka 
+ (b^\L)^{\ka\mu\nu}h_{\mu\nu} + \cdots)
\psb\ga_{\ka}\psi
\equiv b_{\textrm{expt}}^\ka \psb\ga_\ka\psi,
\eeq
where $h_{\mu\nu} \equiv g_{\mu\nu} - \eta_{\mu\nu}$
is the linearized gravitational field.
In a nonrelativistic weak gravity field,
$h_{\mu\nu}$ can be approximated
by $h_{00} \approx -2\ph$,
$h_{0j} \approx 0$,
and $h_{jk} \approx -2\ph \de_{jk}$,
where $\ph$ is the gravitational potential.
The actual $b^\ka$ coefficients measured in experiments
should be the effective value
\beq
b_{\textrm{expt}}^\ka = b_\eff^\ka 
+ (b^\L)^{\ka\mu\nu}h_{\mu\nu} + \cdots 
\approx b_\eff^\ka 
- 2(b^\L)^{\ka\Sigma\Sigma}\ph,
\eeq
where $\Si\Si$ in the index means
a summation over space and time indices 
in the Sun-centered frame,\cite{datatables}
i.e.,
$(b^\L)^{\ka\Sigma\Sigma}
= (b^\L)^{\ka TT} + (b^\L)^{\ka XX}
+ (b^\L)^{\ka YY} + (b^\L)^{\ka ZZ}$.
We see that this effective coefficient 
depends on the gravitational potential.
Moreover,
the combination $(b^\L)^{\ka\Sigma\Sigma}$
can be constrained
by comparing experiments measuring $b^\ka$
at different elevations.

As an example,
an experiment in Seattle
constrained the $\bt^X_e$ coefficient,
a combination of $b^\ka$ 
and other SME coefficients
in the electron sector,\cite{datatables}
as $|\bt^X_e|<3.7\times10^{-31}$ GeV.\cite{brh08}
Another experiment in Taiwan
measured the same combination and got
$|\bt^X_e|<3.1\times10^{-29}$ GeV.\cite{lsh03}
The results are obtained at different elevations
with different gravitational potentials,
so we can compare them
to get a constraint on the linearized coefficient
$\bt_e^{X\Si\Si}$
as $|\bt_e^{X\Si\Si}| < 3.2\times10^{-15}$ GeV.
Similar analyses can be done for other coefficients
using experimental data 
summarized in the data tables\cite{datatables} 
for the SME.\cite{kl21b}
More experiments can be done
to measure SME coefficients in different gravitational potentials,
and those can be used to constrain the linearized coefficients.

\section{Free-fall experiments}

Aside from comparing results in different experiments,
we can also constrain the linearized coefficients
by single gravity-related experiments.
To better analyze those experiments,
we derived the nonrelativistic Hamiltonian 
from the linearized Lagrange density.
The Hamiltonian can be written as
\beq
H=H_0 + H_{\ph} + H_{\si\ph} + H_{g} + H_{\si g} + \ldots,
\eeq
where $H_0$ is the Hamiltonian without background fields,
and other components are the corrections from background fields.
Components with subscript $\si$ are spin dependent,
and those without are spin independent.
The exact terms in the components
can be found in Ref.\ [\refcite{kl21b}].

The Hamiltonian can modify the gravitational acceleration
experienced by a dynamical system
on the Earth's surface.
The spin-dependent components 
permits us to study spin--gravity couplings in the SME framework
for the first time.
In this section,
we use free-fall experiments 
to test the spin--gravity couplings.

One experiment\cite{mgt14} 
compares the effective gravitational accelerations
of two isotopes of strontium atoms,
the spin-zero bosonic ${}^{88}$Sr 
and the spin-9/2 fermionic ${}^{87}$Sr.
Unpolarized ${}^{87}$Sr atoms were used there,
so if effective gravitational accelerations
depend on spin orientations,
the measured gravitational accelerations of ${}^{87}$Sr atoms
should span a broader range than those of ${}^{88}$Sr atoms.
They found no such effect 
to a sensitivity of $10^{-7}$.
Analyzing the result in our framework,
we get bounds on nonrelativistic SME coefficients as
\begin{eqnarray}
\Big|(k^\NR_{\sigma\phi})_n^Z\Big|&<&1\times10^{-4}\textrm{ GeV},\nn\\
\Big|(k^\NR_{\sigma\phi pp})_n^{ZJJ}-0.4(k^\NR_{\sigma\phi pp})_n^{ZZZ}\Big|
&<&5\times10^{-2}\textrm{ GeV}^{-1},
\end{eqnarray}
where the subscript $n$ means the coefficients are for neutrons,
and repeated $J$ indices mean a summation over 
special coordinates $J=X,Y,Z$ in the Sun-centered frame.

Another experiment\cite{xcd16}
compares the gravitational acceleration experienced 
by ${}^{87}$Rb atoms with different spin orientations.
They found no difference
to a sensitivity of $10^{-7}$. 
This can be translated to constraints 
on nonrelativistic coefficients as
\bea
\Big|(k^\NR_{\si\ph})_p^Z-0.6(k^\NR_{\si\ph})_e^Z\Big|
&<&2\times10^{-5}\textrm{ GeV},\nn\\
\Big|(k^\NR_{\si\ph pp})_p^{ZJJ}
+0.3(k^\NR_{\si\ph pp})_p^{JJZ}\Big|
&<&7\times10^{-3} \textrm{ GeV}^{-1},
\eea
where the subscripts $p$ and $e$ 
mean proton and electron flavor, respectively.

Another type of interesting free-fall experiment
is to compare the falls of hydrogen H 
and antihydrogen $\overline{\textrm{H}}$.
This can provide insights on CPT symmetry.
Several groups have been designing experiments 
for that.\cite{antihydrogen}
A detailed theoretical analysis of the falls in our framework
can be found in Ref.~[\refcite{kl21b}].
We expect new results from those experiments in the near future.

\section{Gravitational interferometer experiments}

Our nonrelativistic Hamiltonian 
can also modify the gravity-induced phase shift
in gravitational interferometer experiments.
In this section,
we analyze several interferometer experiments with neutrons
and use them to extract bounds on the nonrelativistic coefficients.

The first gravitational interferometer experiment
was performed by Colella, Overhauser, and Werner (COW).\cite{cow75}
They used Bragg diffraction to split a coherent neutron beam
into two paths at different heights
and measured the relative gravity-induced phase shift 
between the two paths.
Unpolarized neutron beams were used in the experiment,
so it is mainly sensitive 
to the spin-independent terms in our Hamiltonian.

The effective gravitational acceleration
measured in the original COW experiment
attains an accuracy of 10\%.
From this,
we deduce a constraint on a nonrelativistic coefficient as 
\beq
(k^\NR_\ph)_n<1\times10^{-1}\textrm{ GeV},
\eeq
where $(k^\NR_\ph)_n$ is a spin-independent coefficient
in the neutron sector.
More recent versions of the COW experiment
can improve this result.\cite{kl21b}

The next type of interferometer experiments we consider
is the OffSpec experiment,
which uses polarized nonrelativistic neutron beams
and splits the beams by magnetic fields.\cite{voh14}
This is sensitive to spin--gravity couplings.
The experiment measured the effective gravitational acceleration 
to an accuracy of 2.5\%.
After some analysis,\cite{kl21b}
we get the constraint 
\beq
\Big|(k^\NR_\ph)_n+(k^\NR_{\si\ph})^j_n \hat{s}^j\Big|
<2.5\times10^{-2}\textrm{ GeV},
\eeq
where $(k^\NR_\ph)_n$ and $(k^\NR_{\si\ph})^j_n$
are coefficients in the neutron sector,
and $\hat{s}^j$ is the initial polarization direction 
of the neutron beams.
We expect a more precise result to be obtained 
from a more detailed analysis of the experiment.
Also,
our understanding of spin--gravity couplings 
and Lorentz violation
can be further improved by future experiments 
using similar setups with the OffSpec experiment.
For example,
the coefficients $(k^\NR_{\si g})^{jk}_n$ in our Hamiltonian
can be constrained by comparing the phase shifts
between horizontally split neutron beams
with different spin orientations.

\section{Gravitational bound-state experiments}

Another application of the nonrelativistic Hamiltonian 
concerns gravitational bound-state experiments,\cite{vn02,gc18}
where the bounds states of neutrons
in the Earth's gravitational field
are measured.
In those experiments,
our nonrelativistic Hamiltonian
can modify the energy states
by changing the potential experienced by the neutrons.
Specifically,
the spin-independent terms in the Hamiltonian
shift the energy levels,
and the spin-dependent terms
split the energy levels.

The first gravitation bound-state experiment\cite{vn02}
measured the critical heights of the bound states,
which are related to the energy levels.
The precision of the measurement is around 10\%.
A later experiment\cite{gc18}
improved the precision to around 0.3\%
by measuring the transition frequencies 
between different energy levels.
From those results,
constraints on nonrelativistic SME coefficients 
are found to be\cite{kl21b,iwb21}
\bea
\big|(k^\NR_\ph)_n\big|&<&1\times10^{-3}\textrm{ GeV},
\nn\\
\sqrt{\big[(k^\NR_{\si\ph})_n^J\big]^2}&<&8\times10^{-3}\textrm{ GeV},
\eea
where  $(k^\NR_\ph)_n$ and $(k^{\NR}_{\si\ph})_n^J$
are nonrelativistic coefficients in the neutron sector,
and the square implies a summation over $J=X,Y,Z$
in the Sun-centered frame.
We expect these results to be improved by more precise future measurements.

\section*{Acknowledgments}

This work was supported in part by the U.S.\ Department of Energy 
and by the Indiana University Center for Spacetime Symmetries (IUCSS).

\end{document}